\documentclass[twocolumn,showpacs,preprintnumbers]{revtex4}
\usepackage{graphicx}
\usepackage{dcolumn}
\usepackage{bm}

\begin{document}

\preprint{APS/123-QED}
\title{Irreversible Incremental Behavior in a Granular Material}
\author{Luigi La Ragione$^{1,}$}
\author{Vanessa Magnanimo$^{1}$}
\author{James T. Jenkins$^{2}$}
\author{Hernan A. Makse$^{3}$}
\affiliation{$^{1}$Dipartimento di Ingegneria Civile e Ambientale\\ Politecnico di Bari, Via Orabona 4, 70125 Bari, Italy\\ $^{2}$
Department of Theoretical and Applied Mechanics\\ Cornell University, Ithaca, NY 14853 U.S.A. \\ $^{3}$Levich Institute and Physics
Department \\ CCNY, New York, NY 10031 U.S.A.\\ }
\date{\today }

\begin{abstract}
We test the elasticity of granular aggregates using increments of  shear and volume strain in a numerical simulation. We find that
the increment in volume strain is almost reversible, but the increment in shear strain is not. The strength of this irreversibility
increases as the average number of contacts per particle (the coordination number) decreases. For increments of volume strain, an
elastic model that includes both average and fluctuating motions between contacting particles reproduces well the numerical results
over the entire range of coordination numbers. For increments of shear strain, the theory and simulations agree quite well for high
values of the coordination number.

\end{abstract}

 \maketitle

Granular materials have received the attention of many researchers in the last decade because of unsolved problems with direct
relevance to chemistry, physics and engineering. The behavior of a granular material can range between that of a gas and that of a
solid, depending on the applied loading and the regime of deformation considered. Significant progresses have been made with the
introduction of numerical tools (e.g. \cite{Cundall.Strack.1979}) that permit the detailed analysis of an aggregate of particles.
Although such simulations have provided information about inter-particle interactions, such as their elasticity, sliding, and
deletion, and statistical measures of their cooperative behavior, such as induced anisotropy and force chains, it is still unclear
how to incorporate these informations in a predictive theoretical model.

Attempts to do this have been made in the context of the effective medium theory (EMT) (e.g. \cite{Digby, Walton}) in which the
contact displacements are given by the applied average strain. However, the predicted shear and bulk moduli are far from those
measured in numerical simulations (e.g. \cite {Cundall.Jenkins.1989, Makse.Gland.1999}). In order to improve the theoretical
prediction, particle displacements are given by the sum of an average and the fluctuation components (e.g. \cite{Koend, Misra}). In
particular, better predictions of the shear and bulk moduli have been obtained in a recent work that employs pair fluctuations
\cite{Luigi, LaRaJen}. However, Agnolin and Roux \cite{Ivana} point out that such predictions fail when the the coordination
number, $\bar{Z}$, is low and close to the isostatic limit $\bar{Z}_{iso}$ at which the aggregate is statically determinate. They
suggest that for low coordination number, more complicated models are needed that account for collective deformations among
particles. That is, the assumption of pair fluctuations is not sufficient to capture the response of poorly coordinate aggregates
of particles. Here, we address this issue and provide an alternative interpretation of the failure of an elastic description at low
coordination numbers.

We carry out numerical simulations using a distinct element method and focus on the first incremental response of an isotropically
compressed random aggregate that consists of identical, elastic, frictional spheres. We consider dense aggregates, with solid
volume fractions $\phi$ near $0.64$, that have different $\bar{Z}$. Previous work \cite {Ivana, Ell, vanHecke} have
considered poorly coordinated aggregates and found that the shear modulus is proportional to $\bar{Z}-\bar{Z}_{iso}$, where
$\bar{Z}_{iso}$ is equal to four for a packing of frictional spheres. Here, we find that when $\bar{Z}$ decreases, there is an
irreversible behavior of the aggregate that involves local, coordinated, irreversible motions of the particles that are not
resisted by forces. These motions result in a reduction of the apparent stiffness of the aggregate (e.g. \cite{Mouk}). That is, the
initial configuration of a poorly coordinated aggregate can not sustain any incremental strain unless a change in the geometry of
the packing occurs. When such irreversible changes are present, simple elastic theory can not reproduce the response of the
aggregate and the utility of the elastic moduli is questionable.

We introduce measures of these irreversible deformations which vary with the coordination number. For increments in volume strain,
the number of irreversible motions is so small that their effect is negligible and the response of the aggregate can be assumed to
be elastic. More importantly, for shear increments, the strength of the irreversibility persists even at high $\bar{Z}$ and
increases as $\bar{Z}$ decreases towards its isostatic value, indicating that elasticity does not describe the aggregate response.

Our numerical simulations consider $10,000$ particles, each with diameter $d$
$=0.2$ $mm$, randomly generated in a periodic cubic cell. We employ material properties typical of glass spheres: a shear modulus
$\mu=29 $ $GPa$ and a Poisson's ratio, $\nu =0.2$. The interaction between particles is a non-central contact force in which the
normal component is the non-linear Hertz interaction and the tangential component is bilinear: an initial elastic displacement
followed by Coulomb sliding (e.g.\cite{Makse.Gland.1999}). We create different initial isotropic states by varying the coefficient
of friction between particles during the preparation (e.g. \cite{Magna}); all initial states have a solid volume fraction $\phi\sim
0.64$ and a confining pressure that varies from $50 KPa$ to $10 MPa.$

For all packings, we evaluate the response of the aggregate to
homogeneous increments in volume and shear strain by setting the
particle coefficient of friction high enough to prevent sliding. Because of the random positions of the particles and their different initial contact stiffnesses, the subsequent particle motions are the sum of the homogeneous applied strain and a fluctuation that relaxes the particles towards a new equilibrium state. When this relaxation
involves a rearrangement of particles, the incremental response is
irreversible.

\begin{figure}[!tbp]
\resizebox{1.0\columnwidth}{!}{   \includegraphics{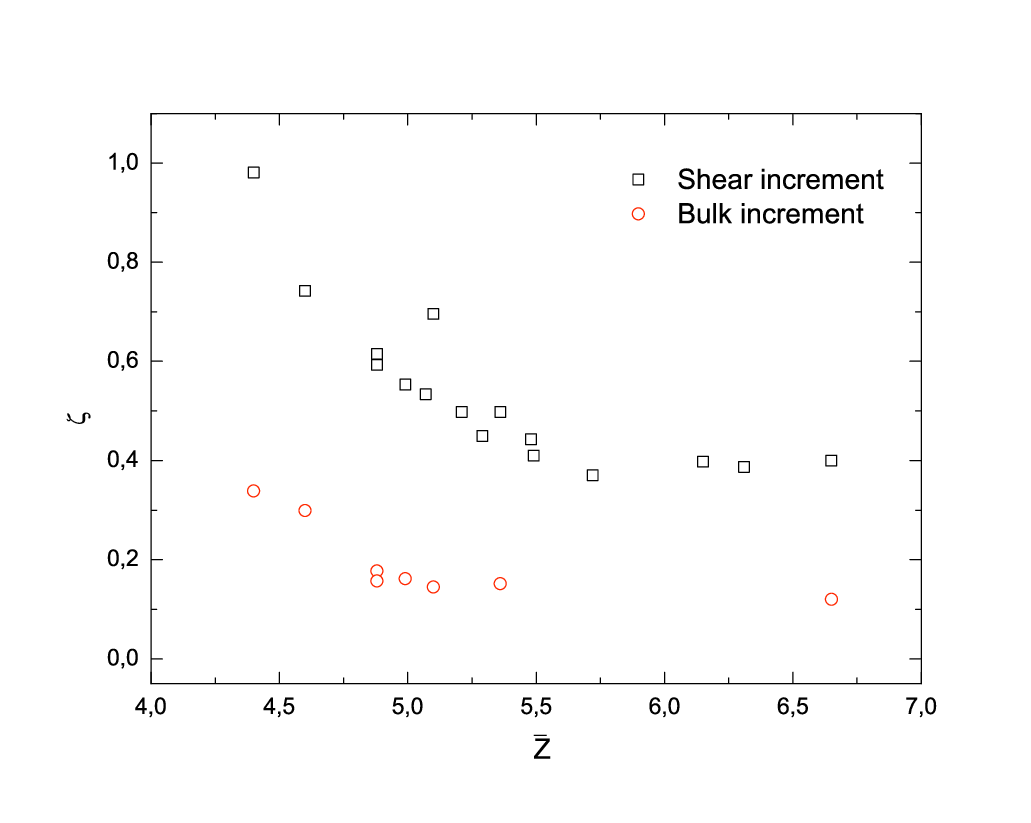} } \vspace{ -0.8cm} \caption{Measurement of irreversibility in
terms of displacements when increments of volume and shear strain are applied.} \label{anom}
\end{figure}

\begin{figure}[!tbp]
\resizebox{1.0\columnwidth}{!}{   \includegraphics{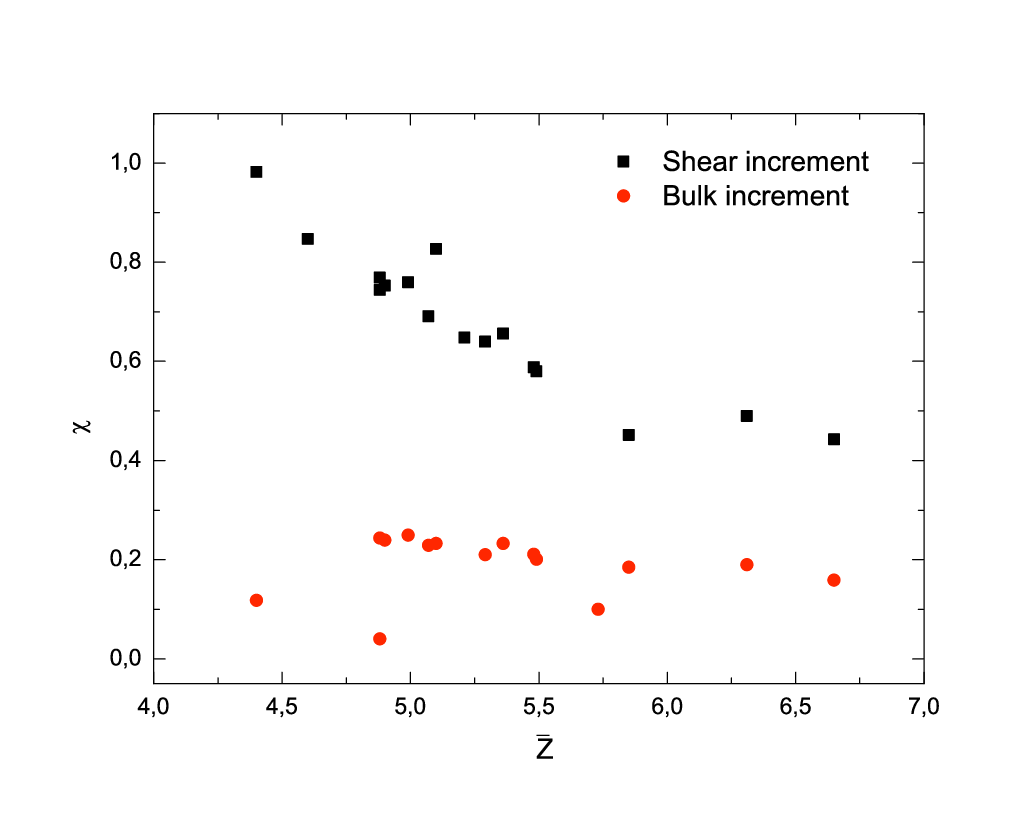} } \vspace{ -0.8cm} \caption{Measurement of irreversibility
in terms of contact forces when
 increments of volume and shear strain are applied. } \label{anom_2}
\end{figure}

We attempt to characterize the particle rearrangement and introduce two measures of the strength of the irreversibility: the first
is related to contact displacement, and the second is associated with contact forces. We apply a forward increment in strain
followed by an identical backward strain in which the reference configuration should be recovered if the deformation were perfectly
elastic.

We define the parameter $\zeta$ as the average over all contacts of the ratio of the absolute values of the contact displacements
after and before the backward increment in strain. A parameter, $\chi$, is similarly defined in terms of the contact forces. Both
parameters would be zero for perfectly elastic behavior.

We measure $\zeta $ and $\chi $ in all of the packings for increments in both shear and volume strain. We apply increments of
strain with magnitudes that depend on the confining pressure; the ratio of the associated volume strain with the confining pressure
is constant at about $10^{-2}$. The results are plotted in Figs. \ref{anom} and \ref{anom_2}, displaying $\zeta $ and $\chi $ as
functions of $\bar{Z}$.

Figs. \ref{anom} and \ref{anom_2} show small variations in $\zeta$ and $\chi$ associated with increments in volume strain. The
strength of the irreversibility is almost negligible, with slight increases as $\bar{Z}_{iso}$ is approached. We conclude that an
approximate elastic response for the aggregate is obtained for increments of volume strain, independent of the coordination
number.

For increments in shear strain, both $\zeta $ and $\chi $ increase as $\bar{Z}_{iso}$ is approached. We believe that these
irreversible motions are associated with the presence of local instability (e.g. \cite{Mouk}).  Moreover, both $\zeta $ and $\chi $
are non-zero for high values of the coordination number, in contrast to what we find for increments in volume strain. The aggregate
seems to experience rearrangements over the entire range of the coordination number, with the irriversibility becoming stronger as
isostaticity is approached.

The condition of isostaticity has been an object of great interest for many researchers (\cite{Makse.Gland.1999, Wyart, Silbert,
Ell, vanHecke, brito}). The contact forces in an aggregate are uniquely determined in terms of the applied loads, independent of
the contact stiffness, when the aggregate is both statically and kinematically determinate (e.g. \cite{Pelle}). The condition for
static determinacy, often referred to as Maxwell's condition \cite{Max}, insures the equality between the number of unknowns and
the number of equilibrium equations. In a granular aggregate, this necessitates that $\bar{Z}=\bar{Z}_{iso} $. The condition for
kinematic determinacy insures that there are no inextensional mechanisms in the aggregate; then a rigid aggregate is able to
sustain any external self-equilibrated perturbation without changing the relative positions of its particle centers.

The kinematic condition is not often taken into account in recent work on granular aggregates (e.g. \cite{Wyart2}) and sometimes
has been emphasized in a different way. For example, Moukarzel \cite{Mouk}, in his description of network rigidity, defines an
isostatic system to be one in which the rank K of the rigidity matrix that relates the contact forces to the applied forces always
equals to the number of equations, rather than one in which the simple Maxwell condition is satisfied. Then, when K is less than
the number of equations, the network is flexible; this can occur when $\bar{Z}=\bar{Z}_{iso} $.

The presence of irreversibility in all of our aggregates, even in the limit that $\bar{Z}=4$ (see Figs. \ref{anom} and
\ref{anom_2}) indicates that the kinematic condition does not hold. Consequently, inextensional mechanisms and the associated soft
or floppy vibrational modes are possible. This situation may occur whatever the value $\bar{Z}$, if particles location do not
correspond to those of a rigid network. In particular, in packing characterized as isostatic by $\bar{Z}=\bar{Z}_{iso} $ (e.g.
\cite{Wyart, Silbert, Ell}),  soft modes can occur as long as the kinematic condition is not satisfied. However, we should emphasize
that the initial states that we employ are constructed in a much
different way than those constructed to insure an initial, stable, elastic response (e.g. \cite{O'Hern}).

We now turn from irreversibility to elasticity and report results from numerical simulation for the bulk modulus, $\overline{\Theta
} =\displaystyle \frac {1}{3} \displaystyle\sum_{i=1}^{3}\Delta \sigma _{ii}/\Delta V$, and the shear modulus, $\overline{G}=\Delta
\sigma _{12}/\Delta \varepsilon _{12}$, where $ \sigma _{ij}$ is the average stress tensor, $\varepsilon _{ij}$ the average strain
tensor, and $V$ is the volume of the aggregate. When a shear increment is applied to poorly coordinated systems, the
irreversibility increases and the material response deviates from elasticity. That is, an elastic theory can only be considered as
providing an upper bound on the shear modulus.

To make predictions of the moduli, we adopt the fluctuation theory developed by \cite{LaRaJen}. This theory improves upon the
simplest average strain models \cite{Digby, Walton}, because contacting particles are assumed to move with both the average
deformation and fluctuations and because the statistics of the aggregate are taken in account. The theory employs force and moment
equilibrium for a typical pair of particles to evaluate the fluctuations, and then uses them to determine the stress in the
aggregate. However, at low values of the coordination number, the simple statistical model introduced to describe the variability
of the neighborhood of contacting pairs of particles is probably too simple, (e.g. \cite{Agn_08}) and the initial distribution
contact forces (e.g.\cite{TA}) is not taken into account.

Here we repair the second deficiency and assume that the distribution, $w(P),$ of the normal component of the contact force, $P$ is
exponential (e.g.\cite{Cundall.Jenkins.1989}):

\begin{equation}
w(P)=\frac{1}{\bar{P}}\exp \left( -\frac{P}{\bar{P}}\right) , \label{density}
\end{equation}%
where, by definition, $\bar{P}=\displaystyle\int_{0}^{\infty }Pw\left( P\right) dP.$

From the effective medium theory (e.g.\cite{Jen-Strack}), the confining pressure $p_{0}$ can be expressed as function of $\bar{P}$,
$$p_{0}=\displaystyle\bar{Z}\phi\bar{P}/ \pi d^{2},$$ and the relation between the normal component of the contact displacement,
$\delta ,$ and $\bar{P}$ is
\begin{equation}
\delta =\left[ \frac{3\left( 1-\nu \right) }{2\mu d^{1/2}}\bar{P}\right] ^{2/3}.  \label{norm_aver}
\end{equation}

Using (\ref{density}) in (\ref{norm_aver}), we obtain

\begin{eqnarray}
\overline{\delta ^{1/2}}&=&\left[ \frac{3\left( 1-\nu \right) }{2\mu d^{1/2}} \right] ^{1/3}\int_{0}^{\infty }P^{1/3}w\left(
P\right) dP \nonumber\\ &=&\left[ \frac{3\left( 1-\nu \right) }{2\mu d^{1/2}} \right] ^{1/3}\bar{P}^{1/3}\Gamma \left(
\frac{4}{3}\right) ,
\end{eqnarray}

where $\Gamma $ is the Gamma function.

So the average normal contact stiffness $\overline{K_{N}}=\mu d^{1/2}\overline{\delta ^{1/2}}/\left( 1-\nu \right)$ becomes
\begin{equation}
\overline{K_{N}}=d\left( \frac{1}{3}\right) ^{1/2}\left[ \frac{9\sqrt{3}\pi \mu ^{2}}{2\overline{Z}\left( 1-\nu \right) ^{2}\phi
}p_{0}\right] ^{1/3}\Gamma \left( \frac{4}{3}\right). \label{KN}
\end{equation}

The average shear stiffness is $\overline{K_{T}}=2\overline{K_{N}}\left( 1-\nu \right) /\left( 2-\nu \right)$.

Taking in account the initial distribution of forces in this way, we obtain a new solution for the fluctuations; with this, the
resulting expressions for the bulk modulus $\overline{\Theta }$ and the shear modulus $\overline{G}$ are, respectively,

\begin{figure}[!tbp]\includegraphics [width=1.0\columnwidth]{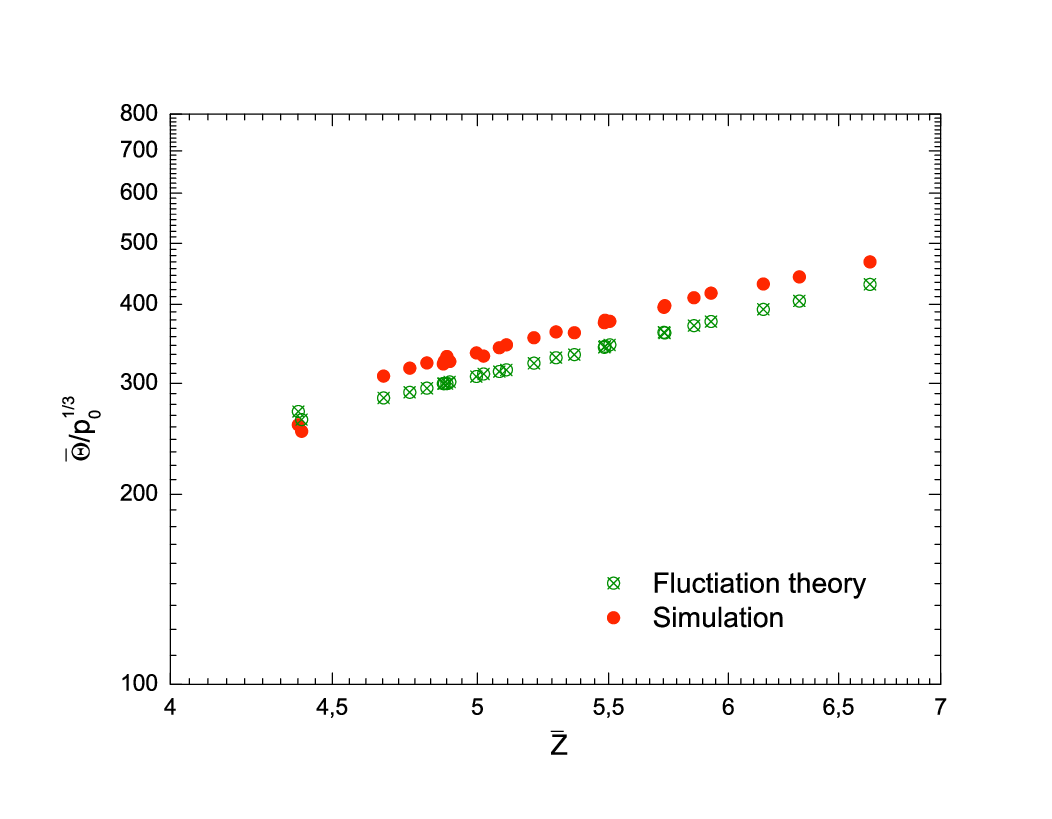}
\caption{Comparison between the numerical data and fluctuation
theory for the normalized bulk modulus.} \vspace{-0.5cm}
\label{comparison_1}
\end{figure}

\begin{figure}[!tbp]
\includegraphics [width=1.0\columnwidth]{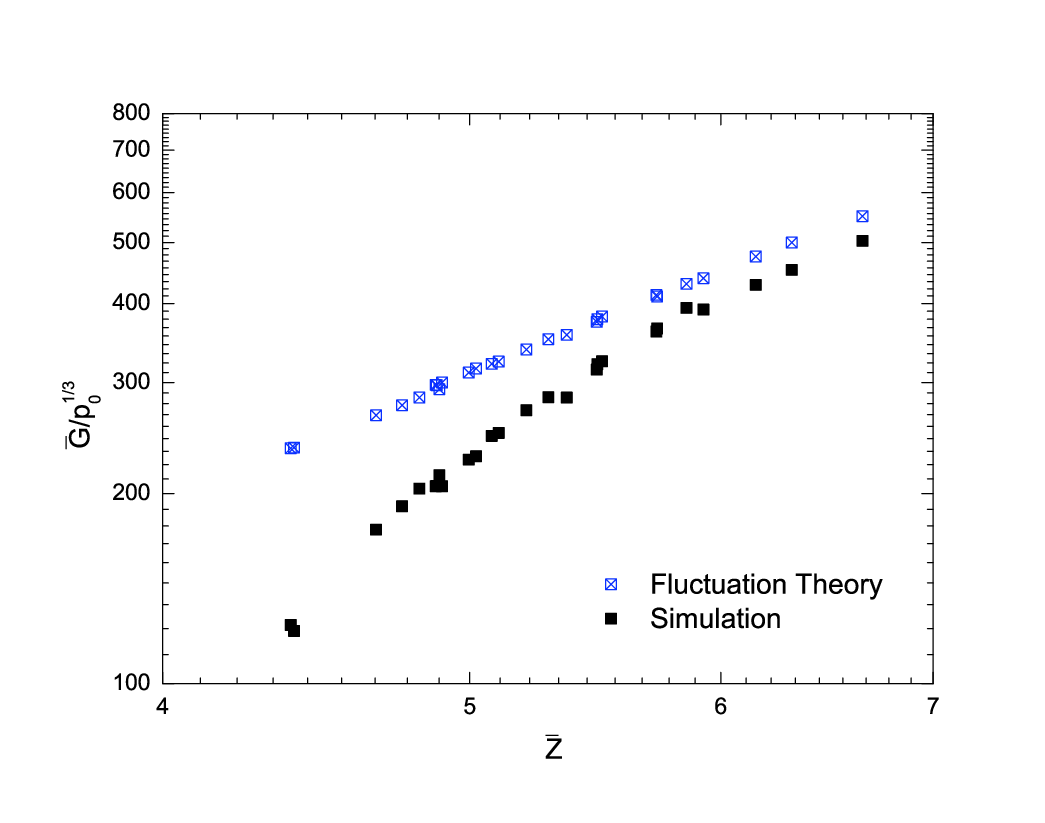}
\caption{Comparison between the numerical data and fluctuation
theory for the normalized shear modulus.} \label{comparison_2}
\end{figure}

\begin{equation}
\overline{\Theta }=\frac{\phi\overline{K_{N}}}{5\pi d} \left[ -2.8\bar{Z} - \frac{14.5}{\bar{Z}^{3}} + \frac{38}{\bar{Z}^{2}} -
\frac{33.6}{\bar{Z}} -12.7 \right] \label{Bulk_F}
\end{equation}

\begin{eqnarray}
\hspace{-0.4cm} \overline{G} &=&\frac{\phi \left(\overline{K_{N}}-\overline{ K_{T}}\right) }{5\pi d}\left[
1.7\bar{Z}+\frac{8.7}{\bar{Z}^{3}}-\frac{22.8}{ \bar{Z}^{2}}+\frac{20.2}{\bar{Z}}-7.6\right]   \nonumber  \label{Sh_G} \\
&&+\frac{\phi \overline{K_{T}}}{5\pi d}\left[ 6.6\bar{Z}+\frac{66.9}{\bar{Z}
^{3}}-\frac{154.7}{\bar{Z}^{2}}+\frac{128.6}{\bar{Z}}-46\right] .
\end{eqnarray}%
where a relation between the rms value of the fluctuation in the number of contacts per particle and its average value has been
adopted \cite{Magna}. At the upper limit of validity of equations (\ref{Bulk_F},\ref{Sh_G}), $ \overline{Z}=22/3$, we recover the
average strain prediction \cite{Digby, Walton}. This limit results from the modeling of the statistical distribution of particles
in the assembly.

Comparison between the predictions of the bulk and shear moduli and
measurements in the numerical simulation are shown in Figs.
\ref{comparison_1} and \ref{comparison_2}. There is agreement at
high values of the coordination number for the shear modulus; the
slight difference may be attributed to the irreversibility. However,
when the comparison is extended to lower values of $\overline{Z},$
the prediction strongly deviate from the simulations, as also seen
by others \cite{Ivana}. Here, we conclude that this discrepancy is
due to the observed irreversible motions and that an elastic theory
is not able to capture the behavior of the system for low values of
$\overline{Z}$. For the bulk modulus, the theory works quite well
over the entire range of $\overline{Z}$, because the
irreversibility, measured by $\chi$ and $\zeta$, can be neglected.

In conclusion, we have measured the irreversible motions in a granular aggregate subjected to incremental strains and determined
their influence on the mechanical response of the material. We found that particles may experience rearrangements in their geometry
and contact forces, even when small perturbations are applied and that these rearrangements are sensitive to the coordination
number. The strength of this irreversibility is negligible for increments in volume strain, while it strongly increases for
increments in shear strain as $\overline{Z}$ approaches  $\overline{Z}_{iso}$. The presence of irreversible motions in the
aggregates indicates a deviation from elastic behavior. That is, an elastic theory is appropriate to describe the material behavior
only if mechanisms do not play an important role in the displacements of the particles.

La Ragione, Magnanimo, and Jenkins are grateful for support from Strategic Plan-119, Regione Puglia (Italy) and Gruppo Nazionale
della Fisica Matematica. Makse acknowledges support from DOE and NSF.

\end{document}